\documentclass[preprint,12pt]{elsarticle}

\makeatletter
\def\ps@pprintTitle{%
 \let\@oddhead\@empty
 \let\@evenhead\@empty
 \def\@oddfoot{\centerline{\thepage}}%
 \let\@evenfoot\@oddfoot}
\makeatother

\usepackage{amssymb}
\usepackage{color}
\usepackage{amsmath}
\usepackage{fullpage}
\usepackage{booktabs,caption}
\usepackage[flushleft]{threeparttable}
\usepackage{multirow}
\usepackage{graphicx}
\usepackage{subcaption}
\usepackage{epstopdf}
\graphicspath{ {figures/} }

\def\nn{\nonumber}

\def\sinh{\textrm{sinh}}

\usepackage{makecell}
\def\correspondingauthor{\footnote{Corresponding author.}}

\begin{document}

\title{Stochastic Local Volatility models and the Wei-Norman factorization method}

\author[unjaen]{Julio Guerrero } 
\address[unjaen]{University of Jaen - Department of Applied Mathematics, Campus Las Lagunillas, Jaen, E-23071, jguerrer@ujaen.es}

\author[uniba]{Giuseppe Orlando\correspondingauthor{}} 
\address[uniba]{University of Bari - Department of Economics and Finance, Via C. Rosalba 53, Bari, I-70124, giuseppe.orlando@uniba.it}

\begin{abstract}
In this paper, we show that a time-dependent local stochastic volatility (SLV) model can be reduced to a system of autonomous PDEs that can be solved using the Heat kernel, by means of the Wei-Norman factorization method and Lie algebraic techniques. Then, we compare the results of traditional Monte Carlo simulations with the explicit solutions obtained by said techniques. 
This approach is new in the literature and, in addition to reducing a non-autonomous problem into an autonomous one, allows for reduced time in numerical computations. 
\end{abstract}
\begin{keyword} Wei-Norman, Lie algebraic methods, Stochastic local volatility (SLV) model\\

\MSC[2010]  60Gxx, 35C05, 35K05, 17Bxx, 62P05
\end{keyword}

\maketitle

\section{Introduction}

Stochastic local volatility (SLV) models lie in the conjunction of stochastic volatility and local volatility models. Stochastic volatility denotes the case in which the volatility of the underlying process is stochastic. Well known examples of those models are Hull and White (1987)\cite{Hull1987}, Stein and Stein (1991) \cite{Stein1991}, Heston (1993) \cite{Heston1993}, etc.

The economic reason behind stochastic volatility models is linked to: a) the fact that equity of a firm can be seen "as the net present value of all its future income plus its assets minus its debt. These constituents have very different relative volatilities which gives rise to a leverage related skew" \cite{Jackel2004}; b) higher demand for downwards risk insurance because of intrinsic asymmetry of positions in equity (which is mostly held long than short); c) asymmetric effects of changes in prices on portfolio rebalancing (e.g. stop-loss condition and regulatory limits). 

Local volatility models were initially developed for daily traded options as the arbitrage-free forward volatilities were inferred and 'locked-in' by market quotes \cite{Alexander2008}. Early examples can be found in Dupire (1994) \cite{Dupire1994}, Derman and Kani (1994) \cite{Derman1994} and Rubinstein (1994) \cite{Rubinstein1994}. Subsequently, by "local volatility" has been indicated any deterministic volatility model in which forward volatilities are a function of both time and the underlying.

Both, stochastic volatility and local volatility models compete for explaining from different angles the instantaneous volatility of the underlying which is unobservable. However, in terms of pricing and hedging contingent claims the two frameworks yield identical results \cite{Alexander2008}.
A difference between the two approaches lies in the way in which the absence of static or dynamic arbitrage is demonstrated. However, while for the stochastic (implied) volatility
framework the no-arbitrage condition requires specific constraints on the state probability space that are difficult to meet \cite{Carr2005} \cite{Davis2007}, "the derivation of no-arbitrage conditions in a stochastic local volatility framework is relatively straightforward because the static no-arbitrage condition is simply that local volatilities are non-negative and the admissible state space is relatively simple" \cite{Alexander2008}.

As mentioned, SLV models provide a general framework combining practical features from stochastic volatility (SV) and local volatility (LV) models \cite{Cui2018}. In this respect, one of the most prominent models is the Stochastic Alpha Beta Rho (SABR) model introduced by Hagan et al. (2002) \cite{Hagan2002}, because provides closed-form asymptotic implied volatility formulas. In addition, the SABR model leads to "fast and accurate pricing, hedging, and calibration for European-style derivatives with moderate duration" and 
predicts the "co-movements between smile dynamics and the underlying
(a forward price or forward rate), which improves upon previous models" \cite{Cui2018}. 
Notwithstanding the many appealing features, SLV models pose a major challenge when it comes to the simulation of the two stochastic processes of the state variables due to the nonlinear local volatility structure.

To date the literature on the SLV models falls into either in Monte Carlo methods (e.g. Chen et al. (2012) \cite{Chen2012}, Leitao et al. (2017) \cite{Leitao2017}) or in asymptotic probability density approximations (e.g. Hagan et al. (2002) \cite{Hagan2002}, Armstrong et al. (2017) \cite{Armstrong2017}, Barger and Lorig (2017) \cite{Barger2017}).\\ 

In this work, we contribute to the literature by using Lie algebraic methods to reduce a time-dependent SLV model to a system of autonomous PDEs that can be solved using appropriate kernels that can be derived from the usual Heat kernel.
 Then, we take the explicit solutions obtained by Lie algebraic techniques and compare them with Monte Carlo simulations. The advantage of this approach lies in reducing a non-autonomous problem into various autonomous ones and in saving numerical calculations. Finally, we show how the obtained prices can be used for deriving the implied volatility.\\

The paper is organized as follows: Section \ref{Sec:SLV} describes the general stochastic local
volatility framework. Section  \ref{Sec:WeiNorman} provides the Wei-Norman solutions.  Sections  \ref{Sec:Kernel-c0} and \ref{Sec:Kernel-c1} the detailed computations of the kernels when $c=0$ and $c=1$ respectively. Sections  \ref{Sec:Price-c0} computes the Price function for the case $c=0$ and  Section \ref{Sec:NumSim} reports the simulations  obtained with the Wie-Norman approach versus those obtained with a classical Monte Carlo. 
Section \ref{Sec:Conclusion} concludes.

\section{SLV model} \label{Sec:SLV}

Local volatility (LV) models, first proposed by  Dupire (1994) \cite{Dupire1994} and Derman et al. (1996) \cite{Derman1996}, are of the type 
\begin{equation}
  dS_t = \mu S_t dt + \sigma_{LV} (S_t, t) S_t dW_t \label{Eq:LV}
\end{equation}
 where $S_t$ is the price of the underlying, $\mu$ the drift, $\sigma_{LV}$ the volatility, and $W_t$ a Wiener process. 

The advantage of these models is that, in an arbitrage-free setting, they provide a perfect fit to the implied volatility surface through Dupire's formula. 
The disadvantage of LV models is that the implied volatility shape soon flattens out as the maturity increases. Therefore, the longer the maturity, the less the resemblance to the near-term smile. 

Stochastic volatility (SV) models are not affected by that problem and they are of the type
\begin{equation}
 \left\{ 
 \begin{array}{lll}
  dS_t &=& S_t \mu dt + m(v_t) S_t dW_t^{(1)}\\
  dv_t &=& \mu(v_t)dt + \sigma(v_t) dW_t^{(2)}
 \end{array}
\right. \label{Eq:SV}
\end{equation}
 with $\mathbb{E}[dW_t^{(1)} dW_t^{(2)} ] = \rho\: dt$, where $W_t^{(1)}$ and $ dW_t^{(2)}$  are two correlated Wiener processes through the correlation coefficient $\rho \in (-1, 1)$.

Given the fact that the implied volatility surface generated from such models generally preserves the near-term smile shape, future smiles and term structure are more realistic but they are difficult to calibrate \cite{Dai2016}. \\

Stochastic Alpha Beta Rho (SABR) model was first introduced by Hagan et al. (2002) \cite{Hagan2002} when they observed that market smiles and skews obtained by local volatility models produced a market smile opposite of observed market behaviour i.e. "when the price of the underlying decreases, local vol models predict that the smile shifts to higher prices; when the price increases, these models predict that the smile shifts to lower prices" \cite{Hagan2002}.
The SABR model, as initially proposed, is 
\begin{equation}
 \left\{ 
 \begin{array}{lll}
  dS_t &=& v_t S_t^{\beta} dW_t^{(1)}\\
  dv_t &=& \alpha \, v_t  dW_t^{(2)}
 \end{array}
\right. \label{Eq:SABRHagan}
\end{equation}
 where $\alpha$ and $\beta$ are some parameters and, as well as with the stochastic volatility framework, the two Wiener processes are linked through the correlation coefficient $\rho$. However as the "stochastic alpha beta rho model for interest rate derivatives was designed for an environment of 5\% base rates", "its traditional implementation method based on a lognormal volatility expansion breaks down in today’s low-rate and high-volatility environment, returning nonsensical negative probabilities and arbitrage" \cite{Balland2013}. For this reason a number of variants have been proposed to date (e.g. Doust (2012) \cite{Doust2012}, Balland et al. (2013)  \cite{Balland2013}, Van der Stoep (2014) \cite{VanderStoep2014}, etc.).

Stochastic local volatility (SLV) models combine desirable features from both stochastic volatility models and local volatility models and enable a closer fit to the
volatility surface of term structures (for an introduction see \cite{Jackel2004}\cite{Diavatopoulos2021}). SLV models are a generalization of the SABR ones and they are expressed as 
%
%
\begin{equation}
 \left\{ 
 \begin{array}{lll}
  dS_t &=& \omega(S_t, v_t )dt + m(v_t )\Gamma(S_t)dW_t^{(1)}\\
  dv_t &=& \mu(v_t)dt + \sigma(v_t)dW_t^{(2)}
 \end{array}
\right. \label{SLVeq}
\end{equation}
 with $\mathbb{E}[dW_t^{(1)} dW_t^{(2)} ] = \rho\: dt$, where $\rho \in (-1, 1)$.

Define the pricing operator $P_t\Phi(S, v) = \mathbb{E}[\Phi(S_t, v_t)|S_0 = S, v_0 = v]$.
The family $(P_t)$ is determined by its infinitesimal generator ${\cal L}^S$, where
\begin{equation}
 {\cal L}^S\Phi(S, v) := \lim_{t\rightarrow 0^+} \frac{P_t\Phi(S, v) -\Phi(S, v)}{t} \label{def_generator_SLV}
\end{equation}

From Eq. (\ref{SLVeq}) it can be shown that \cite{Cui2018}:
\begin{equation}
 {\cal L}^S \Phi= \frac{(m(v)\Gamma(S))^2}{2} \frac{\partial^2\Phi}{\partial S^2}+\rho\: \Gamma(S)\sigma(v) \frac{\partial^2\Phi}{\partial S\partial v}+\frac{\sigma(v)^2}{2} \frac{\partial^2\Phi}{\partial v^2}+\omega(S,v)\frac{\partial\Phi}{\partial S}+\mu(v)\frac{\partial\Phi}{\partial v}\label{generator_SLV}
\end{equation}

According to \^{I}to's lemmas, any function $\Phi(S_t,v_t)$  of the stochastic variables $S_t$ and $v_t$ verifying  the two-dimensional
time-homogeneous diffusion system (\ref{SLVeq})  satisfies the following PDE:
 \begin{equation}
 \frac{\partial\Phi}{\partial t}= {\cal L}^S \Phi \label{PDE}
 \end{equation}

 In the following, we shall work on the setting of the PDE (\ref{PDE}), rather than on the stochastic Eq. (\ref{SLVeq}).
 
 
 To solve the PDE (\ref{PDE})
we shall use the Wei-Norman factorization method  
 \cite{WeiNorman1963, WeiNorman1964}, which allows solving this kind of PDEs even in the case of time-dependent coefficients.
  The price to be paid by such a method is that it can be only applied when ${\cal L}_t^{S} $ is a linear combination of the generators of a Lie algebra of constant differential operators, and this will impose severe restrictions on the functions appearing in ${\cal L}^{S}$. 
More precisely, we shall solve the generalized time-dependent PDE
 \begin{equation}
 \frac{\partial\Phi}{\partial t}= {\cal L}_t^S \Phi \label{timedependent_PDE}
 \end{equation}
 with
 \begin{equation}
   {\cal L}_t^S= \alpha_1(t)K_1 + \alpha_2(t)K_2+\alpha_3(t)K_3+\alpha_4(t)K_4+\alpha_5(t)K_5
 \end{equation}
 where the differential operators $K_i,\,i=1,\ldots,5$ are given by:
 \begin{eqnarray}
  K_1&=&  \frac{(m(v)\Gamma(S))^2}{2} \frac{\partial^2\ }{\partial S^2} \nn \\
  K_2 &=& \rho\: \Gamma(S)\sigma(v) \frac{\partial^2\ }{\partial S\partial v} \nn \\
  K_3&=& \frac{\sigma(v)^2}{2} \frac{\partial^2\ }{\partial v^2} \label{generators}\\
  K_4&=&\omega(S)\frac{\partial\ }{\partial S} \nn \\
  K_5&=& \mu(v)\frac{\partial\ }{\partial v} \nn
 \end{eqnarray}

These operators close a finite-dimensional Lie algebra (and therefore the Wei-Norman method can be applied) provided that
\begin{eqnarray}
 m(v)&=&m_0  \nonumber\\
 \omega(S,v)&=& k_\omega \Gamma(S)\nonumber \\
 \mu(v)&=&k_\mu \sigma(v) \label{WN-conditions}\\
 \sigma(v)&=&\sigma_0+\sigma_1 v \nn\\
 m_0^2\:\Gamma(S)\: \Gamma''(S) +c &=&0\nonumber
\end{eqnarray}
where $m_0,k_\omega,k_\mu$ and $c$ are constants. The possible values for the constant $c$ are essentially $c=0$ and $c=\pm 1$. Let us discuss the three cases separately.

\subsection{Case $c=0$}

In this case the last equation of (\ref{WN-conditions}) gives:
\begin{eqnarray}
 \Gamma(S)&=& \Gamma_0+\Gamma_1 S\nonumber
\end{eqnarray}
where  $\Gamma_0$ and $\Gamma_1$ are constants.
The resulting Lie algebra is Abelian, i.e. all commutators vanish,  $[K_i,K_j]=0$.

\subsection{Case $c=\pm1$}

In this case we have:
\begin{equation}
 \Gamma(S)= \Gamma_1 e^{ -[{\rm Erf}^{-1}(\sqrt{\frac{2 c }{\pi}}\frac{S-\Gamma_0}{m_0 \Gamma_1})]^2}  
\end{equation}
where ${\rm Erf}$ stands for the error function and $\Gamma_0$ and $\Gamma_1$ are constants. Note that when $c=-1$ the ${\rm Erf}$ function changes to the imaginary ${\rm Erfi}$ function (which is also real): ${\rm Erfi}(z)=-i\: {\rm Erf}(i z)$.

The resulting Lie algebra is 
\begin{equation}
 [K_2,K_1]= \frac{c}{2} K_2 \,\qquad
 [K_4,K_1]= \frac{c}{2} K_4  \label{LieAlgc}
\end{equation}
the rest of the commutators being zero. This algebra corresponds to that of two affine algebras with the common dilation generator $K_1$.

\section{Wei-Norman solution} \label{Sec:WeiNorman}

Using the Wei-Norman method, the pricing operator\footnote{We have introduced an arbitrary initial time $t_0$, although it will be fixed to $t_0=0$ in most cases. We also select for convenience a particular order in the factorization.} $P_{(t,t_0)}$ associated to the time-dependent PDE (\ref{timedependent_PDE}) can be computed as\footnote{Order of the factors does not change the final expression
of the pricing operator, but it can facilitate its computation.}:
\begin{equation}
 P_{(t,t_0)} = e^{g_1(t,t_0) K_4} e^{g_2(t,t_0) K_5}e^{g_3(t,t_0) K_2}e^{g_4(t,t_0) K_1}e^{g_5(t,t_0) K_3}
 \label{WNfactor}
\end{equation}
where the functions $g_i(t,t_0),\,i=1,\ldots,4$ verify the linear (in this case) equations:
\begin{eqnarray}
g_1'(t,t_0) &=& \alpha_4(t) -\frac{c}{2}\:\alpha_1(t) g_1(t,t_0) \nn\\ 
g_2'(t,t_0) &=&  \alpha_5(t)  \nn\\
g_3'(t,t_0) &=&  \alpha_2(t) -\frac{c}{2}\:\alpha_1(t) g_3(t,t_0)\\
g_4'(t,t_0) &=&  \alpha_1(t)\nn \\
g_5'(t,t_0) &=&  \alpha_3(t)\nn
\end{eqnarray}
with the initial conditions $g_i(t_0,t_0)=0, \,i=1,\ldots,5$, and the primes indicate derivative with respect to $t$. These equations are valid for the three cases $c=0$ and $c=\pm 1$.

The solutions of these equations are simply given by (just supposing that the $\alpha_i(t),  \,i=1,\ldots,5$ are locally integrable):
\begin{eqnarray}
g_1(t,t_0) &=&  \int_{t_0}^t e^{-\frac{c}{2}\:\int_{t'}^{t}\alpha_1(t'')dt''} \alpha_4(t')dt' \nn\\ 
g_2(t,t_0) &=&  \int_{t_0}^t \alpha_5(t')dt'   \nn\\
g_3(t,t_0) &=& \int_{t_0}^t e^{-\frac{c}{2}\:\int_{t'}^{t}\alpha_1(t'')dt''} \alpha_2(t')dt' \label{WNEqs}\\
g_4(t,t_0) &=&  \int_{t_0}^t \alpha_1(t')dt' \nn \\
g_5(t,t_0) &=&  \int_{t_0}^t \alpha_3(t')dt' \nn
\end{eqnarray}

Note that $g_i(t,t_0)$ can be interpreted as some kind of  averaged time funtions. For that, suppose we fix $t$, for instance $t=T$ (expiration date), then we can write $g_i(T,t_0)=\bar\alpha_i (T-t_0)$, where  $\bar\alpha_i=\frac{1}{T-t_0}g_i(T,t_0)$ are time averaged coefficients. Thus, at $t=T$ we can substitute the time-dependent coefficients $\alpha_i(t)$ by the averaged constant ones $\bar\alpha_i$ (rendering the problem autonomous) and obtain the same results.
For $c=0$ (Abelian Lie algebra) the time average is the standard one, but for $c=\pm 1$ is more convolved, due to the noncommutativity of the Lie algebra.
These should be compared with other approaches to handle time dependent coefficients with time average methods, like in \cite{Hagan2002}.

In the case of constant coeficients $\alpha_i(t)=\alpha_i^0$, equations (\ref{WNEqs}) simplifies to:
\begin{eqnarray}
g_1(t,t_0) &=&  \alpha_4^0\frac{ \left(1-e^{-\frac{1}{2} \alpha_1^0 c (t-t_0)}\right)}{\frac{1}{2}\alpha_1^0 c} \nn\\ 
g_2(t,t_0) &=&  \alpha_5^0(t-t_0)   \nn\\
g_3(t,t_0) &=& \alpha_2^0\frac{ \left(1-e^{-\frac{1}{2} \alpha_1^0 c (t-t_0)}\right)}{\frac{1}{2}\alpha_1^0 c} \label{WNEqscte}\\
g_4(t,t_0) &=& \alpha_4^0(t-t_0) \nn \\
g_5(t,t_0) &=& \alpha_3^0(t-t_0) \nn
\end{eqnarray}

For the case $c=0$ these equations reduce to:
\begin{eqnarray}
g_1(t,t_0) &=&  \alpha_4^0(t-t_0) \nn\\ 
g_2(t,t_0) &=&  \alpha_5^0(t-t_0)   \nn\\
g_3(t,t_0) &=& \alpha_2^0(t-t_0) \label{WNEqscte0}\\
g_4(t,t_0) &=& \alpha_4^0(t-t_0) \nn \\
g_5(t,t_0) &=& \alpha_3^0(t-t_0) \nn
\end{eqnarray}

To have the complete solution to the problem, we need to give an explicit expression for each of the exponentials of the operators $K_i$ appearing in Eq.  (\ref{WNfactor}).

For that purpose we need to solve the time-independent PDEs:
 \begin{equation}
 \frac{\partial\Phi}{\partial t}= K_i \Phi \label{timedependent_PDE_K}\,\qquad i=1,\ldots,5
 \end{equation}
 in terms of integral operators with kernels $k_i(t,t',S,v,S',v')$:
\begin{equation}
 \Phi(S,v,t)=\int_0^\infty\int_0^\infty dS'dv' k_i(t,t',S,v,S',v')\Phi(S',v',t')
\end{equation}

The kernels $k_i(t,t',S,v,S',v')$ are fundamental solutions of Eqns. (\ref{timedependent_PDE_K}), i.e.:
\begin{eqnarray}
 \frac{\partial k_i(t,t',S,v,S',v')}{\partial t}&=& K_i k_i(t,t',S,v,S',v') \\
 \lim_{t'\rightarrow t} k_i(t,t',S,v,S',v')&=&\delta(S-S')\delta(v-v') 
 \end{eqnarray}
for $i=1,\ldots,5$.
 
 Since the operators $K_i$ are independent of time, the kernels $k_i$ depend on the difference $t-t'$: $k_i(t,t',S,v,S',v')=k_i(t-t',S,v,S',v')$.
 
 Then each exponential in Eq.  (\ref{WNfactor}) can be computed iteratively (starting with the rightmost one) as:
\begin{eqnarray}
\Phi(S,v,t)&=&  \int_0^\infty \int_0^\infty dS_1dv_1 k_4(g_1(t,t_0),S,v,S_1,v_1) \times \nn \\
           & &\int_0^\infty\int_0^\infty dS_2dv_2 k_5(g_2(t,t_0),S_1,v_1,S_2,v_2) \times \nn\\
           & &\int_0^\infty\int_0^\infty dS_3dv_3 k_2(g_3(t,t_0),S_2,v_2,S_3,v_3) \times \label{ProductKernels}\\
           & &\int_0^\infty\int_0^\infty dS_4dv_4 k_1(g_3(t,t_0),S_3,v_3,S_4,v_4) \times \nn \\
           & &\int_0^\infty\int_0^\infty dS_5dv_5 k_3(g_5(t,t_0),S_4,v_4,S_5,v_5) \Phi(S_5,v_5,t_0) \nn
\end{eqnarray}

Selecting $\Phi(S,v,t_0)=\delta(S-S')\delta(v-v')$ in eq. (\ref{ProductKernels}), we obtain as $\Phi(S,v,t)$ the expression for the integral kernel (or fundamental solution, see for instance \cite{Friedmann}) $k(t,S,v,S',v')$ of the original PDE  \label{timedependent_PDE} particularized to the Wei-Norman case.

Once we have the general solution in terms of the kernels $k_i,\,i=1,\ldots,5$, we need to obtain their explicit form.

\vspace{1cm}

\section{Explicit computation of the kernels for the case $c=0$} \label{Sec:Kernel-c0}

To compute the kernels, it is helpful to perform the following change of variables:
\begin{eqnarray}
 X_S&=&\log(\Gamma_0+\Gamma_1 S) \nonumber\\
 X_v&=&\log(\sigma_0+\sigma_1 v) \label{variablechange}
\end{eqnarray}

In the new variables, the differential operators (\ref{generators}) are written as:
\begin{eqnarray}
  K_1&=&  \frac{m_0^2 \Gamma_1^2}{2} \left( \frac{\partial^2\ }{\partial X_S^2} -\frac{\partial\ }{\partial X_S}\right)  \nn \\
  K_2 &=& \rho\: \Gamma_1\sigma_1 \frac{\partial^2\ }{\partial X_S\partial X_v} \nn \\
  K_3&=& \frac{\sigma_1^2}{2}\left( \frac{\partial^2\ }{\partial X_v^2}-\frac{\partial\ }{\partial X_v}\right) \label{generatorstrans} \\
  K_4&=&k_\omega \Gamma_1\frac{\partial\ }{\partial X_S}\nn \\
  K_5&=& k_\mu \sigma_1 \frac{\partial\ }{\partial X_v}\nn 
 \end{eqnarray}

Using the results of \ref{App:Kernel}, we can easily compute the expressions of the kernels for the differential operators (\ref{generatorstrans}), resulting in:
\begin{eqnarray}
 k_1(t,X_S,X_v,X_S',X_v')&=&\frac{1}{\sqrt{2\pi} m_0\Gamma_1 \sqrt{t}} e^{-\frac{1}{8}m_0^2 \Gamma_1^2 t -\frac{(X_S-X_S')^2}{2m_0^2\Gamma_1^2 t} + \frac{1}{2} (X_S-X_S')} \delta(X_v-X_v')  \nn \\
 k_2(t,X_S,X_v,X_S',X_v')&= &\frac{1}{\sqrt{2\pi}\rho\Gamma_1 \sigma_1 t } e^{-\frac{(X_S-X_S')(X_v-X_v')}{\rho\Gamma_1 \sigma_1 t }}\nn \\
 k_3(t,X_S,X_v,X_S',X_v')&=&\frac{1}{\sqrt{2\pi} \sigma_1 \sqrt{t}} e^{-\frac{1}{8} \sigma_1^2 t -\frac{(X_v-X_v')^2}{2\sigma_1^2 t} + \frac{1}{2} (X_v-X_v')} \delta(X_S-X_S')  \\
 k_4(t,X_S,X_v,X_S',X_v')&=&\delta(X_S-X_S'+k_\omega \Gamma_1 t)\delta(X_v-X_v') \nn\\
 k_5(t,X_S,X_v,X_S',X_v')&= &\delta(X_v-X_v'+k_\mu \sigma_1 t)\delta(X_S-X_S') \nn
\end{eqnarray}

It should be stressed that kernel $k_2$ is divergent since it corresponds to a mixed derivative operator. However, when integrated
jointly with kernels $k_1$ and $k_3$ can result in convergent expressions if suitable conditions are satisfied (see later).

The expression of the kernel (or fundamental solution) of the original PDE (\ref{timedependent_PDE}) particularized to the case
$c=0$ is:
\begin{eqnarray}
 k^{(0)}(t,X_S,X_v,X_S',X_v')&=& \frac{1}{\sqrt{2\pi} \Gamma_1\sigma_1\sqrt{\Delta^{(0)}}} e^{-\frac{ \frac{g_5(t,t_0)}{\Gamma_1^2}(X_S-X_S'+k_\omega\Gamma_1 g_1(t,t_0))^2+\frac{m_0^2 g_4(t,t_0)}{\sigma_1^2}(X_v-X_v'+k_\mu\sigma_1 g_2(t,t_0))^2 }{2\Delta^{(0)}}} \nn\\
 & & \times e^{ -\frac{\frac{ \rho g_3(t,t_0) }{2 \sigma_1\Gamma_1 }(X_S-X_S'+k_\omega\Gamma_1 g_1(t,t_0) + m_0^2\Gamma_1^2 g_4(t,t_0))(X_v-X_v' +k_\mu \sigma_1 g_2(t,t_0)+\sigma_1^2 g_5(t,t_0))}{2\Delta^{(0)}}} \label{Kernel0}\\
 & & \times e^{\frac{m_0^2 g_4(t,t_0)\left( (X_S-X_S'+X_v-X_v' + k_\omega\Gamma_1 g_1(t,t_0)+ k_\mu\sigma_1 g_2(t,t_0) +m_0^2 \Gamma_1^2 g_4(t,t_0)  )g_5(t,t_0) + \sigma_1^2 g_5(t,t_0)^2\right) }{2\Delta^{(0)}}} \nn
\end{eqnarray}
with $\Delta^{(0)}= m_0^2 g_4(t,t_0)g_5(t,t_0)-\rho^2 g_3(t,t_0)^2$. 

This kernel is integrable if $\Delta^{(0)}>0$, which imposes a (time-dependent) restriction on the 
possible values of $\rho$. For the simplest case of constant parameters $\alpha_i(t)=1, i=1,\ldots,5$ (and therefore $g_i(t,t_0)=t-t_0$), $\Delta^{(0)}=(m_0^2 -\rho^2)(t-t_0)^2$ and
this restriction reduces to
$|\rho|<m_0$. Eventually, $m_0$ can be absorbed in $\Gamma_1$ and then we obtain the more standard restriction $|\rho|<1$.

Note that the expression of the kernel is exact and valid for time dependent coefficients (making use of equations (\ref{WNEqs})).

\section{Price function for the case $c=0$} \label{Sec:Price-c0}

The Price function is a function of the variables $S_t$ and $v_t$, and therefore  satisfies (using \^Ito's lemmas) the PDE  (\ref{timedependent_PDE}) with initial condition $max(S-K,0)\delta(v-v_0)$, $K$ being the strike price and $v_0$ an initial value for the volatility variable. Integrating this initial condition with the kernel (\ref{Kernel0}) (using the variables (\ref{variablechange})) and assuming $v\approx v_0$ the following (independent of $X_v$) Price function is obtained:
\begin{eqnarray}
V^0(t,t_0,X_S,K)&=&
\frac{e^{-\frac{(2k_\mu g_2(t,t_0)+2 \Gamma_1 \rho g_3(t,t_0)-\sigma_1 g_5(t,t_0))^2}{8 g_5(t,t_0)}}}{2 \sqrt{2 \pi } \Gamma_1 \sigma_1 \sqrt{g_5(t,t_0)}} \left( e^{X_S+\Gamma_1 k_\omega g_1(t,t_0)}  \left(1+\text{Erf}(d_1+d_2)\right) \right. \nn \\
 & & 
 \left. - e^{X_K+\frac{\Gamma_1 \rho g_3(t,to) (2 k_\mu g_2(t,t_0)+\Gamma_1 \rho g_3(t,t_0)-\sigma_1 g_5(t,t_0))}{2 g_5(t,t_0)}} 
 \left(1-\text{Erf}(d_1-d_2)\right)\right)
\end{eqnarray}
where
\begin{eqnarray}
 d_1&=& \frac{g_5(t,t_0) (2 X_K-2 X_S-2 \Gamma_1 k_\omega  g_1(t,t_0)-\Gamma_1 \rho  \sigma_1 g_3(t,t_0))+\Gamma_1 \rho  g_3(t,t_0) (2 k_\mu  g_2(t,t_0)+\Gamma_1 \rho  g_3(t,t_0))}{2 \sqrt{2} \Gamma_1 \sqrt{g_5(t,t_0) \Delta^{(0)}}} \nn \\
 d_2&=& \frac{\Gamma_1}{2} \sqrt{  \frac{\Delta^{(0)}}{2g_5(t,t_0)}}
\end{eqnarray}

A surface plot of the Price function $V$ as a function of $S$ and $K$ for fixed $t=T$ is shown in Figure \ref{Fig2}a.

\section{Explicit computation of the kernels for the case $c=\pm 1$} \label{Sec:Kernel-c1}

As in the case $c=0$, it is helpful to perform a change of variables, which in this case is:
\begin{eqnarray}
 X_S&=& \sqrt{\frac{2}{c}} m_0 \Gamma_1 \textrm{Erf}^{-1}\left(\sqrt{\frac{2c}{\pi}} \frac{S-\Gamma_0}{m_0\Gamma_1}\right) \nonumber\\
 X_v&=&\log(\sigma_0+\sigma_1 v) \label{variablechange-c}
\end{eqnarray}
contrarily to the case $c=0$, in the case $c=\pm 1$ the allowed values of the variable $S$ (and $K$) are restricted to a finite interval 
$S\in [\Gamma_0-\Delta S,\Gamma_0+\Delta S]$, with $\Delta S=\sqrt{\frac{\pi}{2c}}m_0\Gamma_1$. Notice that the said boundary may suit those option's strategies such as butterfly or iron condor where the positive pay-off is limited between two values \cite{Thomsett2018}.

In the new variables, the differential operators (\ref{generators}) have the same expressions as in (\ref{generatorstrans}), except for the case of $K_1$, which now is written as:
\begin{equation}
  K_1 =  \frac{m_0^2 \Gamma_1^2}{2} \left( \frac{\partial^2\ }{\partial X_S^2} -c\frac{X_S}{m_0^2\Gamma_1^2}\frac{\partial\ }{\partial X_S}\right) 
 \end{equation}

 The kernels are also the same (but in the new variables (\ref{variablechange-c})) as for the case $c=0$, except for $k_1$, which now reads:
\begin{equation}
k_1(t,X_S,X_v,X_S',X_v')=\frac{1}{\sqrt{2\pi} m_0\Gamma_1 \sqrt{e^{c t/2}\frac{\sinh (ct/2)}{c/2} }} 
e^{ -\frac{(e^{ct/2}X_S-X_S')^2}{2m_0^2\Gamma_1^2 e^{c t/2} \frac{\sinh (ct/2)}{c/2}} } \delta(X_v-X_v')  
\end{equation}

The expression of the kernel (or fundamental solution) of the original PDE (\ref{timedependent_PDE}) particularized to the case
$c=\pm 1$ is:
\begin{eqnarray}
  k^{c}(t,X_S,X_v,X_S',X_v')& =& \frac{1}{2 \pi  \Gamma_1 \sigma_1 \sqrt{ \Delta^{(c)}} } \nn \\
& & \times e^{-\frac{ \sigma_1^2 g_5(t,t_0) \left(e^{\frac{1}{2}c g_4(t,t_0)}X_S-X_S'\right)^2 +\Gamma_1^2 m_0^2 (X_v-X_v')^2 \tilde{g}_4(t,t_0) }{2 \Gamma_1^2 \sigma_1^2 \Delta^{(c)} }} \nn\\  
 & & \times
   e^{\frac{   k_\mu \rho  \sigma_1^2 \Gamma_1    g_2(t,t_0) g_3(t,t_0) e^{\frac{1}{2} c g_4(t,t_0)} \left(e^{\frac{1}{2} c g_4(t,t_0)} (X_S+\Gamma_1 k_\omega  g_1(t,t_0))-X_S'\right)}{\Gamma_1^2 \sigma_1^2 \Delta^{(c)} } } \nn\\
& & \times  e^{-\frac{ k_\mu\Gamma_1^2 m_0^2\sigma_1 \tilde{g}_4(t,t_0) \left(  \left(2 (X_v- X_v')-\sigma_1^2 g_5(t,t_0)\right) 
+ k_\mu \sigma_1 g_2(t,t_0)^2 \right) }{2 \Gamma_1^2 \sigma_1^2 \Delta^{(c)} }   } \nn \\
&  &
\times  e^{\frac{ \Gamma_1 \rho  \sigma_1 g_3(t,t_0) e^{\frac{1}{2} c g_4(t,t_0)} \left(2( X_v- X_v')-\sigma_1^2 g_5(t,t_0)\right) \left(e^{\frac{1}{2} c g_4(t,t_0)} (X_S+\Gamma_1 k_\omega  g_1(t,t_0))-X_S'\right) }{2 \Gamma_1^2 \sigma_1^2 \Delta^{(c)} }   } 
\nn\\
& &
\times e^{-\frac{\sigma_1^2 g_5(t,t_0) \Gamma_1 e^{\frac{1}{2}c g_4(t,t_0)} \left(8 c k_\omega  g_1(t,t_0) ( e^{\frac{1}{2} c g_4(t,t_0)}X_S-X_S')+\Gamma_1 k_\omega  g_1(t,t_0))+\Gamma_1 m_0^2 e^{\frac{1}{2} c g_4(t,t_0)}\sigma_1^2 g_5(t,t_0)\right)}{8 \Gamma_1^2 \sigma_1^2 \Delta^{(c)} }   } \nn \\
&  &
\times e^{\frac{-
\Gamma_1^2 m_0^2 \sigma_1^2 g_5(t,t_0)^2+4 \Gamma_1^2 m_0^2\sigma_1^2g_5(t,t_0) (X_v-X_v') \tilde{g}_4(t,t_0)}{8 \Gamma_1^2 \sigma_1^2 \Delta^{(c)} }   } \nn
\end{eqnarray}
where $\Delta^{(c)}= m_0^2 g_5(t,t_0)\tilde{g}_4(t,t_0)-\rho ^2 e^{c g_4(t,t_0)}  g_3(t,t_0)^2$ and 
$\tilde{g}_4(t,t_0)=\frac{e^{c g_4(t,t_0)}-1}{c}$.

This kernel is integrable if $\Delta^{(c)}>0$, which imposes a (time-dependent) restriction on the 
possible values of $\rho$. For the simplest case of constant parameters $\alpha_i(t)=1, i=1,\ldots,5$, this restriction reduces to
$|\rho|< m_0 \frac{\sqrt{c(t-t_0) (e^{c (t-t_0)}-1)}}{2 \left(e^{\frac{1}{2} c (t-t_0)}-1\right)}$, from which we recover 
$|\rho|< m_0$ in the limit $c\rightarrow 0$.

The following research will address the cases where $c=\pm 1$.


\newpage

\section{Numerical simulations} \label{Sec:NumSim}
In this section, we are focusing on the case where $c=0$ provides similar results of the SABR, and we leave the other two cases $c=-1$ and $c=1$ to future research. We first run a Monte Carlo simulation to compare the Wei-Norman SLV model with the SABR model and then we show the implied volatility surface. 

The following sections report the main results in terms of Monte Carlo simulations for the asset $S$ (Sec. \ref{Sec:MC-S}) and the local implied volatility  \ref{Sec:LocalImplVol}. For other results such as the comparison between the mean prices between the Wei-Norman SLV and the SABR, the volatility paths, and the sensitivity to the parameters, the reader may refer to \ref{App:NumSim}.

\subsection{Monte Carlo simulations of $S$} \label{Sec:MC-S}
Concerning numerical simulations, we run Monte Carlo on the SABR described in Eq. (\ref{Eq:SABRHagan}) and on the Wei-Norman specification of the SLV model in Eq. \ref{SLVeq}.

\begin{figure}[!ht]
    \centering
\begin{subfigure}[a]{1\textwidth}
   \includegraphics[width=1\linewidth]{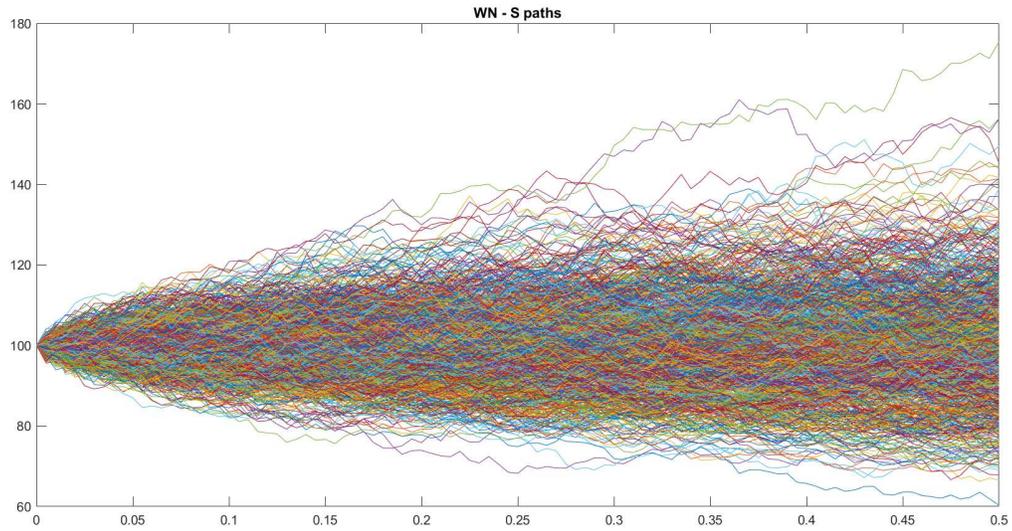}
   \caption{Wei-Norman SLV - Asset ($S$)}
   \label{Fig:WeiNorman-S-c0} 
\end{subfigure}
\begin{subfigure}[b]{1\textwidth}
   \includegraphics[width=1\linewidth]{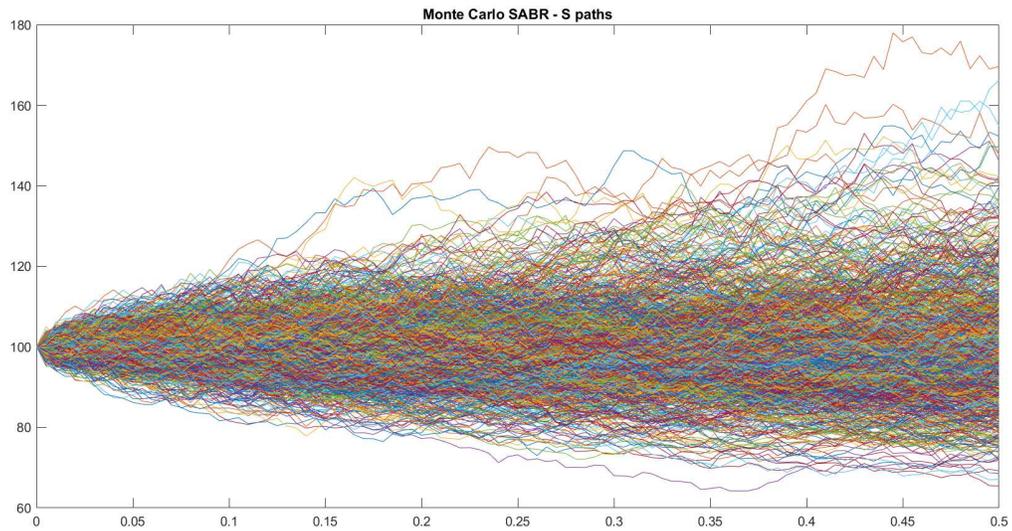}
   \caption{SABR \cite{Hagan2002} - Asset ($S$)}
   \label{Fig:SABR-S-c0} 
\end{subfigure}
\caption{Setting: $S_0=100$, $v_0=0.2$, $\beta=1$, $\rho=0.5$, $\alpha=0.2$. Notice that in absence of dividends and discount factors there is no drift (if we consider the drift coefficients $k_\omega$ and $k_\mu$ zero), therefore, the expected value should be equal to the initial value}
\end{figure}

Table \ref{T:MC-c0} summarizes the results obtained with the Monte Carlo simulations displayed in Figures \ref{Fig:WeiNorman-S-c0} and \ref{Fig:SABR-S-c0} on the asset $S$. As shown, even though the simulations provide similar results, we prove that the Wei-Norman SLV model is more computationally efficient. 

\begin{table}[!ht]
\centering
  \begin{threeparttable}
\caption{Monte Carlo simulations: SABR vs Wei-Norman, case $c=0$}
\begin{tabular}{@{}ccccc@{}}
\toprule
\hline
Monte Carlo & CPU time (sec.) & $S$       & $v$       & $\rho$                \\ \midrule
SABR        & 0.19040         & 0.05000 & 0.19990 & \multirow{2}{*}{1}    \\
Wei-Norman  & 0.10700         & 0.05000 & 0.19980 &                       \\
\hline
SABR        & 0.31250         & 0.04920 & 0.19980 & \multirow{2}{*}{0.5}  \\
Wei-Norman  & 0.18840         & 0.04980 & 0.19790 &                       \\
\hline
SABR        & 0.15460         & 0.04960 & 0.19960 & \multirow{2}{*}{0}    \\
Wei-Norman  & 0.10170         & 0.05000 & 0.20050 &                       \\
\hline
SABR        & 0.14600         & 0.04870 & 0.20040 & \multirow{2}{*}{-0.5} \\
Wei-Norman  & 0.08960         & 0.05010 & 0.19950 &                       \\
\hline
SABR        & 0.16660         & 0.04980 & 0.20010 & \multirow{2}{*}{-1}   \\
Wei-Norman  & 0.12870         & 0.05020 & 0.19800 &                       \\ 
\hline
\bottomrule
\end{tabular} \label{T:MC-c0}
    \begin{tablenotes}
      \footnotesize
      \item Simulations=10,000 for each model; Setting: $T=1$, $S_0=100$, $v_0=0.2$, $\beta=1$, $\alpha=0.2$
          \end{tablenotes}
  \end{threeparttable}
\end{table}

%
%
%
%
%
%
%

\newpage
\pagebreak
\clearpage

\subsection{Implied volatility}  \label{Sec:LocalImplVol}

\begin{figure}[!ht]
    \centering
\begin{subfigure}[a]{1\textwidth}
    \centering
   \includegraphics[width=.65\linewidth]{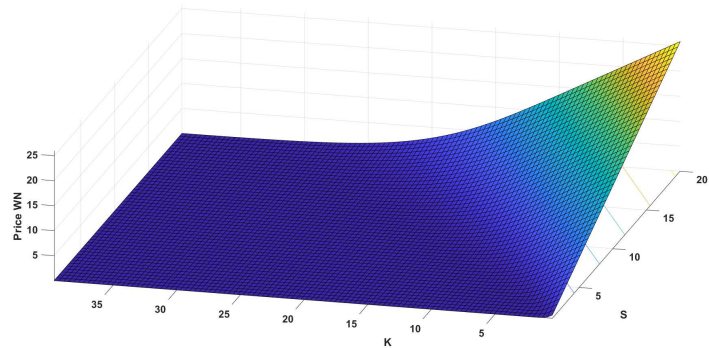}
   \caption{} 
   \label{Fig:WNGridc0} 
\end{subfigure}

\begin{subfigure}[b]{1\textwidth}
    \centering
       \includegraphics[width=.65\linewidth]{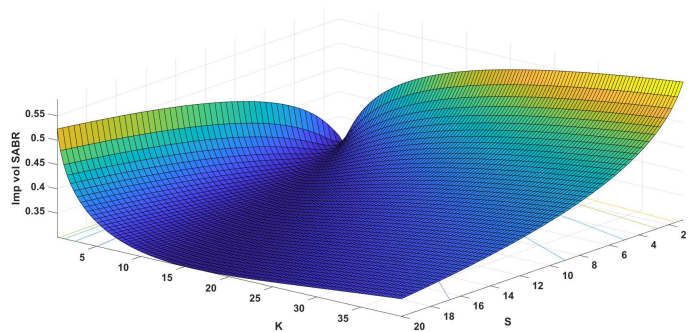}
   \caption{ } 
   \label{Fig:SABRImpliedVolc0} 
\end{subfigure}
  \caption{(a) Wei-Norman SLV prices. (b) Wei-Norman SLV/SABR implied volatility }  
  \label{Fig2}
\end{figure}

Let us remind that the SABR model is very similar to the Wei-Norman SLV when $c=0$. With regard to implied volatility, thanks to the fact that the stochastic volatility process follows a geometric Brownian motion, it is possible to obtain an exact simulation. Figure \ref{Fig:WNGridc0} shows the computed prices and Figure \ref{Fig:SABRImpliedVolc0} the related implied volatility (that, contrarily to the Black and Scholes model, varies according to the moneyness).

Notice that volatility surfaces are characterized by volatility skew, that is why stochastic volatility models can describe well such as phenomenon. The skew might be caused by: a) leverage effect because assets tend to be more volatile at lower prices than at higher prices, b) anti-correlated volatility changes versus spot changes, asymmetric jumps that tend to be downwards rather than upwards, c) a nonzero probability for the asset to collapse if the issuer defaults, d) effect of supply and demand when investors are long and so tend to be net buyers of downside puts and sellers of upside call \cite{Kamal2010}.
This is in line with empirical evidence, in fact, for a given expiration date, implied volatilities increase as strike price decreases for strikes below the current stock price (spot) or current forward price \cite{Kamal2010}. 
Then, the implied volatility surface (as a function of moneyness) is V-shaped but has a rounded vertex when both S and K increase. 

\newpage
\pagebreak
\clearpage

\section{Conclusions} \label{Sec:Conclusion}
In this work, we have shown how a time-dependent local stochastic volatility (SLV) model can be reduced to a system of autonomous PDEs via the Wei-Norman factorization method and  Lie algebraic techniques. For the case $c=0$ the  Wei-Norman SLV is similar to the well-known SABR model with $\beta=1$. By comparing Monte Carlo simulations we have demonstrated that the  Wei-Norman SLV model is more computationally efficient. 
Next research will illustrate the results for $c=-1$ and $c=1$.

\appendix

\section{The Heat kernel and other integral kernels}
\label{App:Kernel}

In this Appendix, we recall some known results for the explicit solutions of some simple PDEs in terms of integral kernels. For an account on this see, for instance, \cite{Friedmann, Bakkaloglu}.

\subsection{The first-order kernel}

Consider the simplest first-order PDE:
\begin{equation}
  \frac{\partial \Phi(x,t)}{\partial t}=  \frac{\partial \Phi(x,t) }{\partial x} 
  \label{1stEq}
\end{equation}

The solution can be expressed in terms of the integral kernel:
\begin{equation}
 k_{\rm 1st}(t,x,x')=\delta(x-x'+t)
 \end{equation}
 which in this case is trivial to integrate:
 \begin{equation}
\Phi(x,t)=\int_{-\infty}^\infty dx' k_{\rm 1st}(t-t_0,x,x')\Phi(x', t_0)=\Phi(x+(t-t_0),t_0)
\end{equation}

\subsection{Transforming a first-order differential operator into a simpler form}

Consider the differential operator:
\begin{equation}
 K=f(x)\frac{\partial\ }{\partial x}\,,\qquad f(x)>0
\end{equation}

This operator can be transformed into  $\frac{\partial\ }{\partial \tilde x}$ by means of the change in the independent variable:
\begin{equation}
 \tilde x=\psi(x)\,,\qquad \psi(x)=\int \frac{dx}{f(x)}
\end{equation}
and then
\begin{equation}
 K=\frac{\partial\ }{\partial \tilde x}
\end{equation}

With this transformation, a general first-order PDE
\begin{equation}
  \frac{\partial \Phi(x,t)}{\partial t}=  f(x) \frac{\partial \Phi(x,t) }{\partial x} \,\,\qquad f(x)>0
  \label{1stfEq}
\end{equation}

can be solved in terms of the kernel
\begin{equation}
 k_{\rm mod1st}(t,x,x')=\delta(\psi^{-1}(\psi(x)+t)-x')\,,
 \end{equation}
 where $\psi^{-1}(x)$ is the inverse function of $\psi(x)$ (which exists since $\psi'(x)=f(x)>0$).
 Thus,
 \begin{equation}
  \Phi(x,t)= \Phi(\psi^{-1}(\psi(x)+(t-t_0)),t_0)
 \end{equation}

\subsection{The Heat Equation}

The Heat equation
\begin{equation}
  \frac{\partial \Phi(x,t)}{\partial t}=  \frac{\partial^2 \Phi(x,t) }{\partial x^2} 
  \label{HeatEq}
\end{equation}
can be formally solved in terms of the integral kernel, known as Heat kernel:
\begin{equation}
 k_H(t,x,x')=\frac{1}{\sqrt{4\pi |t|}} e^{-\frac{(x'-x)^2}{4t}} \label{HeatKernel}
\end{equation}
in the form:
\begin{equation}
\Phi(x,t)=\int_{-\infty}^\infty dx' k_H(t- t_0,x,x')\Phi(x', t_0)
\end{equation}
where $\Phi(x', t_0)$ is some specified function at $t=t_0$. Note that the Heat kernel satisfies:
\begin{equation}
 \lim_{t\rightarrow 0} k_H(t,x,x')=\delta(x-x') 
\end{equation}

\subsection{Transforming a PDE into the Heat equation}

Consider the PDE
\begin{equation}
  \frac{\partial \Phi(x,t)}{\partial t}=  f(x) \frac{\partial^2 \Phi(x,t) }{\partial x^2} \,,\qquad f(x)>0
  \label{PDEf}
\end{equation}
that can be solved using an integral kernel
\begin{equation}
\Phi(x,t)=\int_{-\infty}^\infty dx' k_f(t- t_0,x,x')\Phi(x', t_0)
\end{equation}
where $\Phi(x, t_0)$ is some specified function at $t=t_0$.

We wonder if Eq. (\ref{PDEf}) can be transformed into the Heat equation (\ref{HeatEq}) with a suitable change in the independent and dependent variables:
\begin{equation}
 \tilde{t}=\phi(t) \,,\qquad 
 \tilde{x}= \psi(x,t)\,,\qquad
 \tilde{\Phi}(\tilde{x},\tilde{t})=  \xi(x,t)\Phi(x,t) \label{Changef}
\end{equation}
in such a way that we recover the Heat equation (\ref{HeatEq}) in the new variables:
\begin{equation}
  \frac{\partial\tilde\Phi}{\partial \tilde t}=  \frac{\partial^2\tilde\Phi }{\partial \tilde x^2} 
  \label{HeatEqtilde}
\end{equation}

This is possible if $f$ verifies:
\begin{equation}
 f''(x)-\frac{3}{4}\frac{f'(x)^2}{f(x)}=c_1 g(x)^2 +c_2 \label{Eqf}
\end{equation}
with $c_1,c_2$ constants and $g(x)= \int_0^x\frac{dx'}{\sqrt{f(x')}}$.

In this case, we have that:
\begin{eqnarray}
 \phi(t)&=& \left\{ \begin{array}{ll}
                     t & c_1=0 \\
                     \frac{\tan(\sqrt{c_1}t)}{\sqrt{c_1}} & c_1\neq 0
                    \end{array} \right. \\
 \psi(x,t)&=& \sqrt{\phi'(t)}g(x)\\
 \xi(x,t)&=& e^{-\frac{c_2}{4}t} \left(\phi'(t)f(x)\right)^{-1/4} e^{-\frac{1}{8} \frac{\phi''(t)}{\phi'(t)}g(x)^2}
\end{eqnarray}

Using the Heat kernel (\ref{HeatKernel}), and undoing the change (\ref{Changef}), we obtain the expression for the kernel $k_f$:
\begin{equation}
 k_f(t-t_0,x,x')=\frac{\partial \psi(x',t_0)}{\partial x'}\frac{\xi(x',t_0)}{\xi(x,t)}k_H(\phi(t)-\phi(t_0),\psi(x,t),\psi(x',t_0))
\end{equation}
where it is possible to show that the right-hand side is a function of $t-t_0$ only.

\subsection{The Mixed Equation}

Consider the PDE in 2D:
\begin{equation}
  \frac{\partial \Phi(x,y,t)}{\partial t}=  \frac{\partial^2 \Phi(x,y,t) }{\partial x\partial y} 
  \label{MixedEq}
\end{equation}
which can be formally solved in terms of the integral kernel:
\begin{equation}
 k_M(t,x,y,x',y')=\frac{1}{\sqrt{2\pi} |t|} e^{-\frac{(x'-x)(y'-y)}{t}} \label{MixedKernel}
\end{equation}
in the form:
\begin{equation}
\Phi(x,y,t)=\int_{-\infty}^\infty \int_{-\infty}^\infty dx'dy' k_M(t- t_0,x,y,x',y')\Phi(x',y', t_0)
\end{equation}
where $\Phi(x',y', t_0)$ is some specified function at $t=t_0$. Note that the Heat kernel satisfies:
\begin{equation}
 \lim_{t\rightarrow 0} k_M(t,x,y,x',y')=\delta(x-x')\delta(y-y') 
\end{equation}

\newpage
\pagebreak

\section{Further results of numerical simulations} \label{App:NumSim}

\subsection{Monte Carlo simulations of $v$}  \label{App:MC-Sim-V}

\begin{figure}[!ht]
    \centering
\begin{subfigure}[a]{1\textwidth}
   \includegraphics[width=1\linewidth]{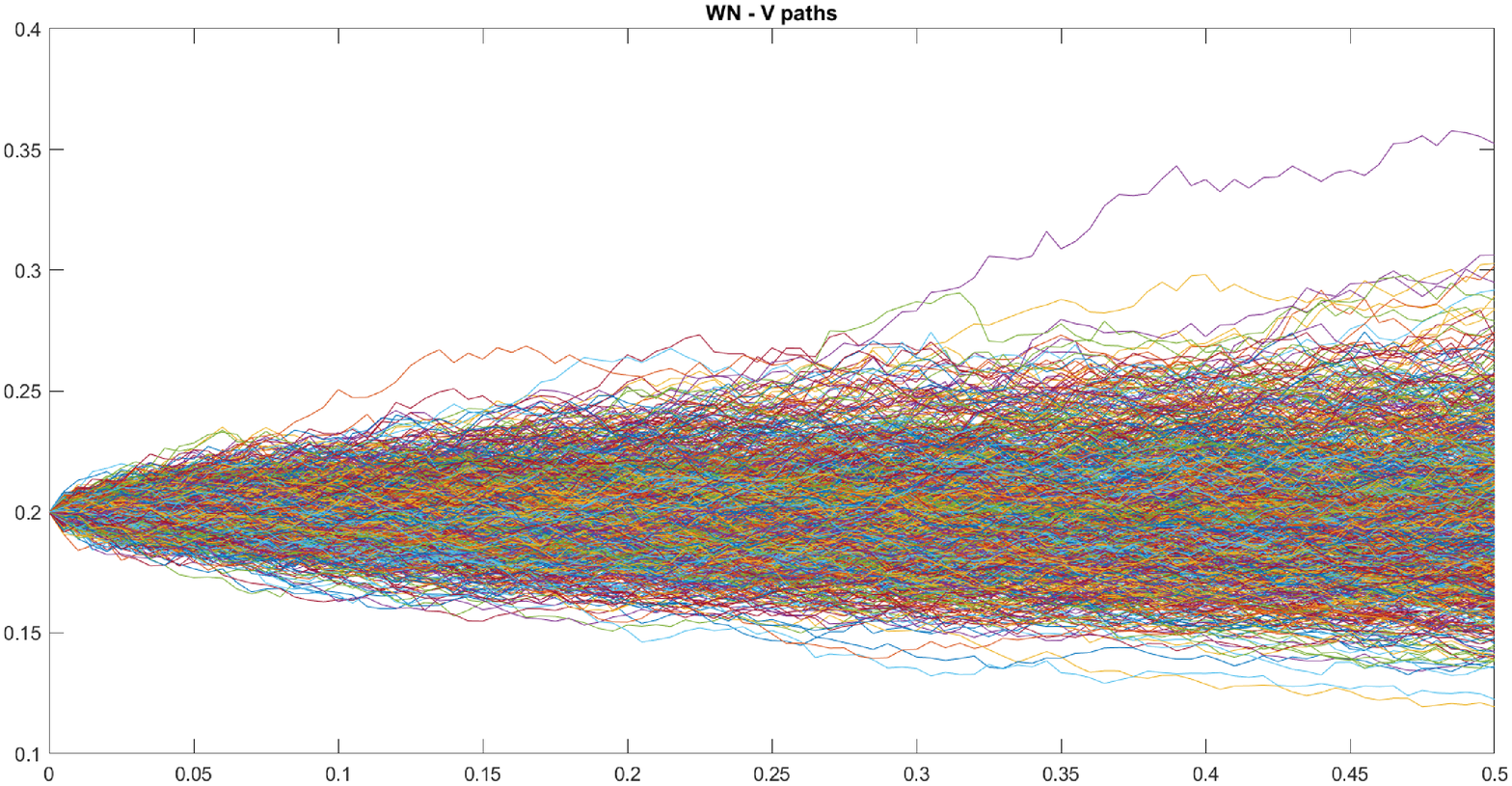}
   \caption{Wei-Norman SLV - Volatility ($v$) }
   \label{Fig:WeiNorman-V-c0} 
\end{subfigure}
\begin{subfigure}[b]{1\textwidth}
   \includegraphics[width=1\linewidth]{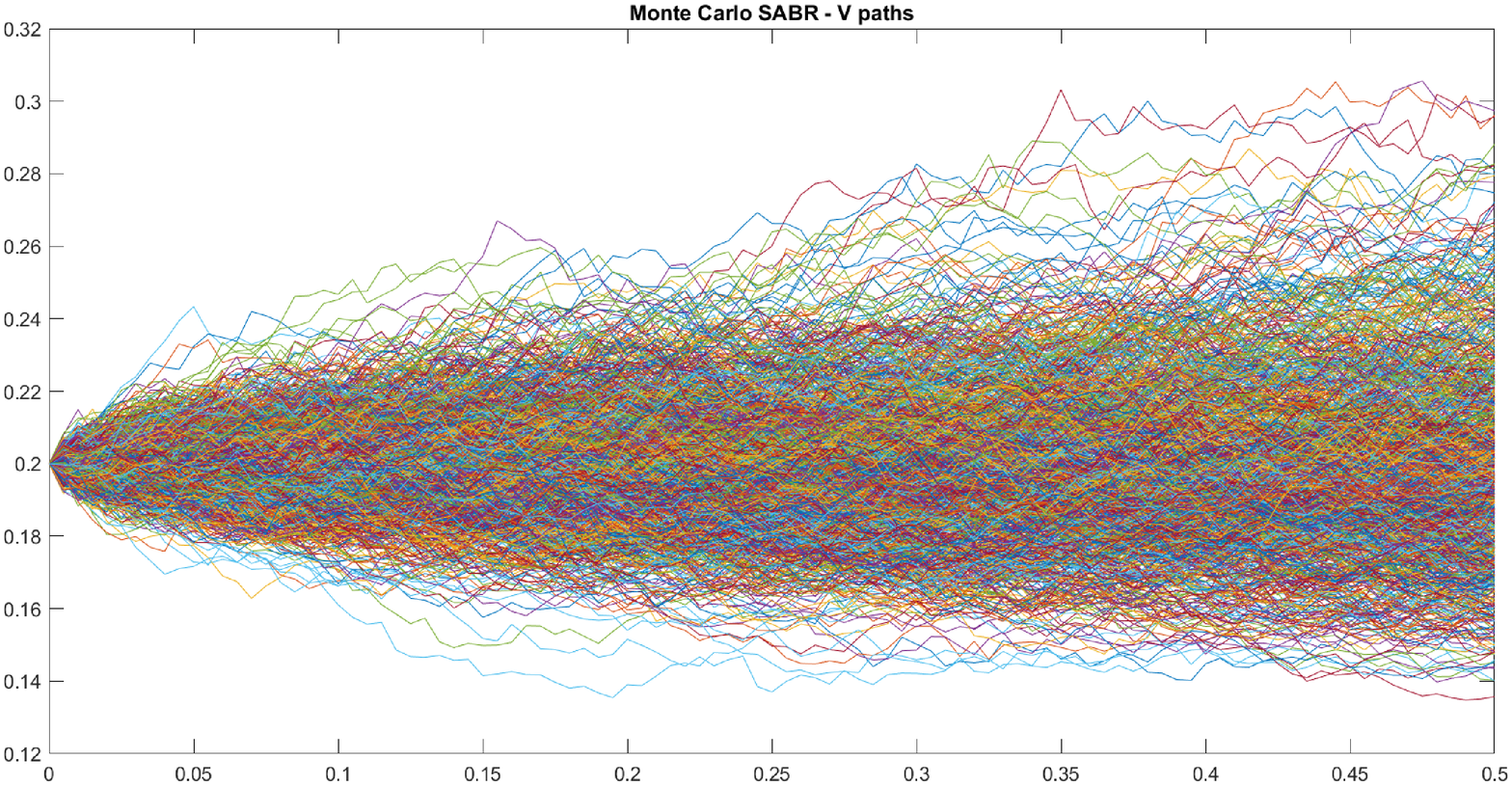}
   \caption{SABR - Volatility ($v$)}
   \label{Fig:SABR-V-c0} 
\end{subfigure}
\end{figure}

Here we show the variable $v$ obtained by running Monte Carlo simulations. Figure \ref{Fig:WeiNorman-V-c0} correspond to the Wei-Norman SLV model and Figure \ref{Fig:SABR-V-c0} correspond to the SABR model. 

\subsection{Standard deviation of Monte Carlo simulations}  \label{App:MC-Sim-Stdev}

\begin{figure}[!ht]
    \centering
\begin{subfigure}[a]{1\textwidth}
   \includegraphics[width=1\linewidth]{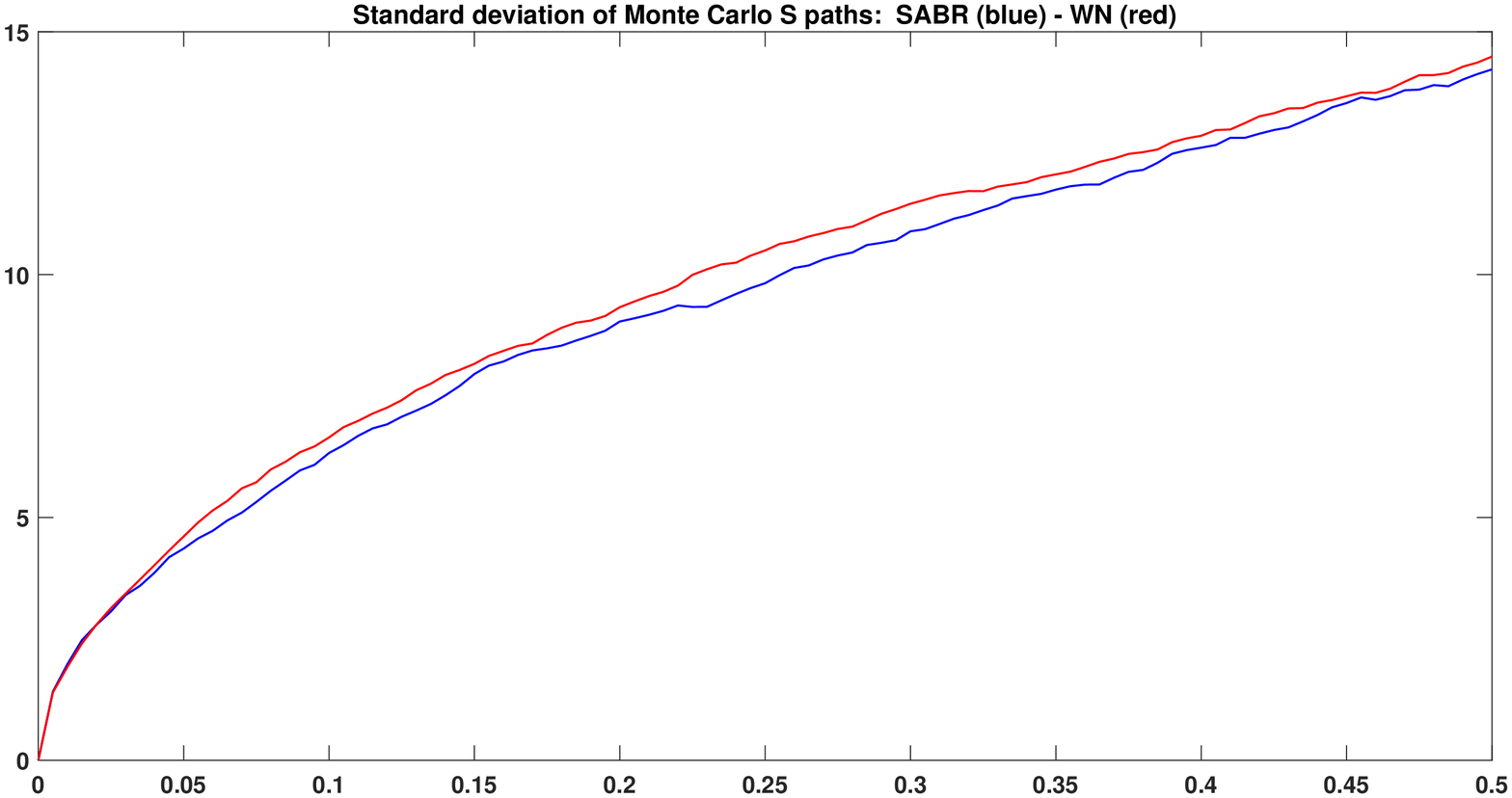}
   \caption{Monte Carlo of Wei-Norman SLV vs SABR - Volatility of $S$ paths}
   \label{Fig:S-StdDev-SABR-WN} 
\end{subfigure}
\begin{subfigure}[b]{1\textwidth}
   \includegraphics[width=1\linewidth]{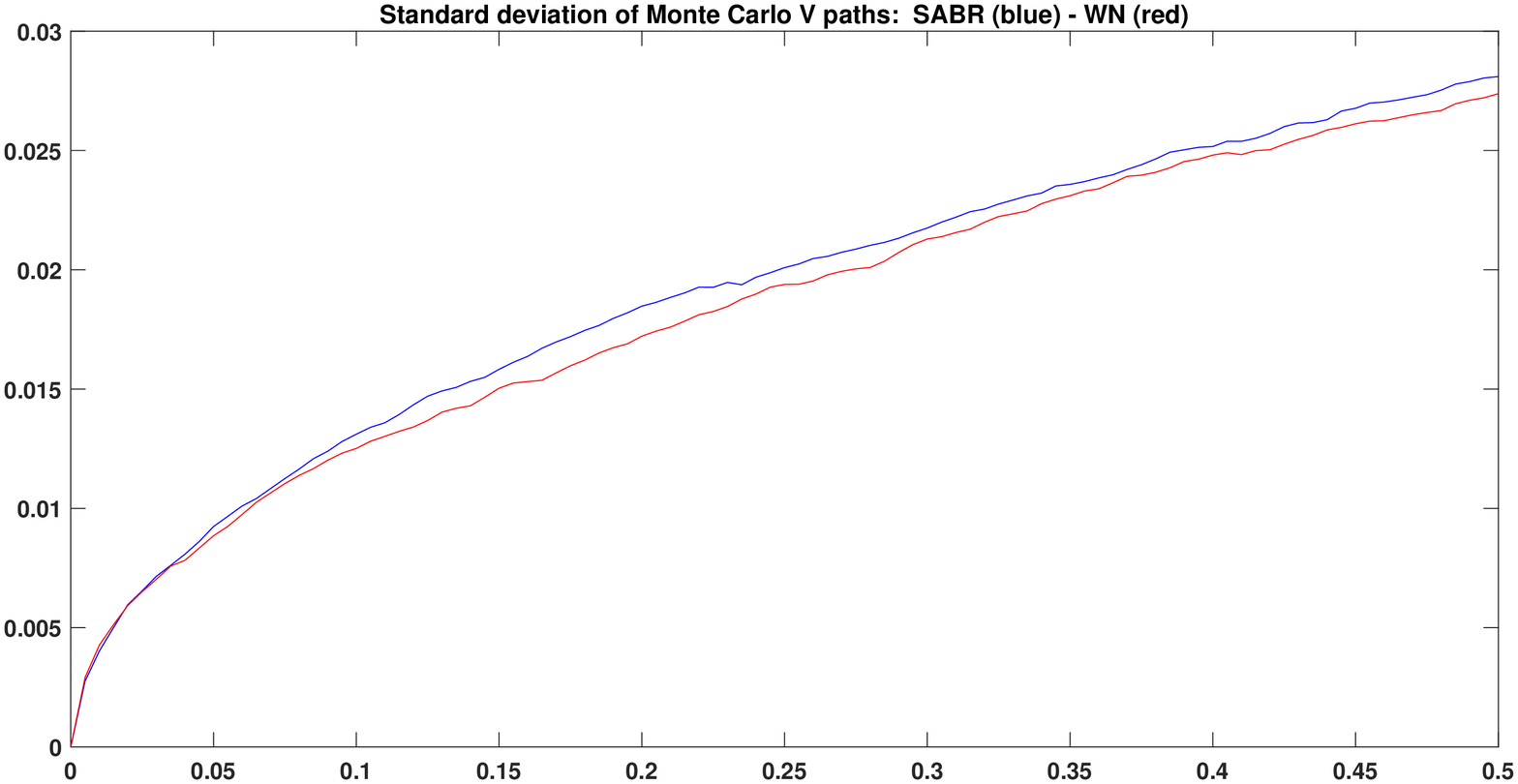}
   \caption{Monte Carlo of Wei-Norman SLV vs SABR - Volatility of $v$ paths}
   \label{Fig:V-StdDev-SABR-WN} 
\end{subfigure}
\end{figure}

Figures \ref{Fig:S-StdDev-SABR-WN} and \ref{Fig:S-StdDev-SABR-WN}, compare the volatility of the obtained Monte Carlo paths in order to confirm that two models produce similar results.

\subsection{Sensitivity to parameters}  \label{Sec:Param}
In the following, we show how the two models perform in case the parameters  $\alpha$ and  $\rho$ are changed. 

\newpage
\pagebreak
\clearpage

\subsection{Sensitivity to $\alpha$}  \label{Sec:ParamAlpha}
This section shows how the two models perform when $\alpha$ changes. Essentially a higher value implies a higher trajectory for both $S$ and $v$. 
\subsubsection{Sensitivity to $\alpha$, asset paths}

To see how a path may change by changing the $\alpha$ Fig. \ref{Fig:WN-V-alpha-c0} and Fig.  \ref{Fig:SABR-V-alpha-c0} show different paths of the asset.
%
%
\begin{figure}[!ht]
    \centering
\begin{subfigure}[a]{1\textwidth}
   \includegraphics[width=1\linewidth]{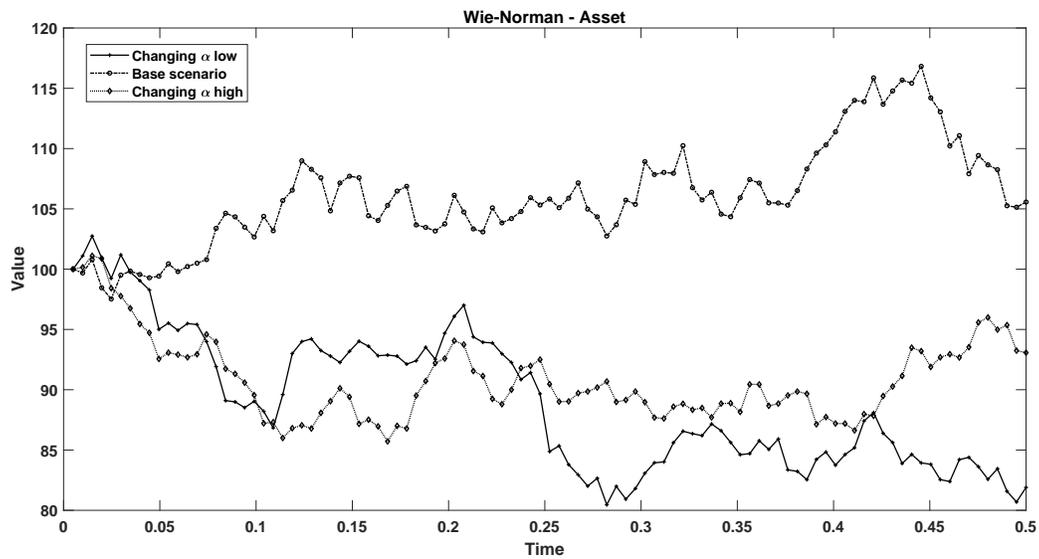}
   \caption{Wei-Norman SLV - Asset}
   \label{Fig:WN-V-alpha-c0} 
\end{subfigure}
\begin{subfigure}[b]{1\textwidth}
   \includegraphics[width=1\linewidth]{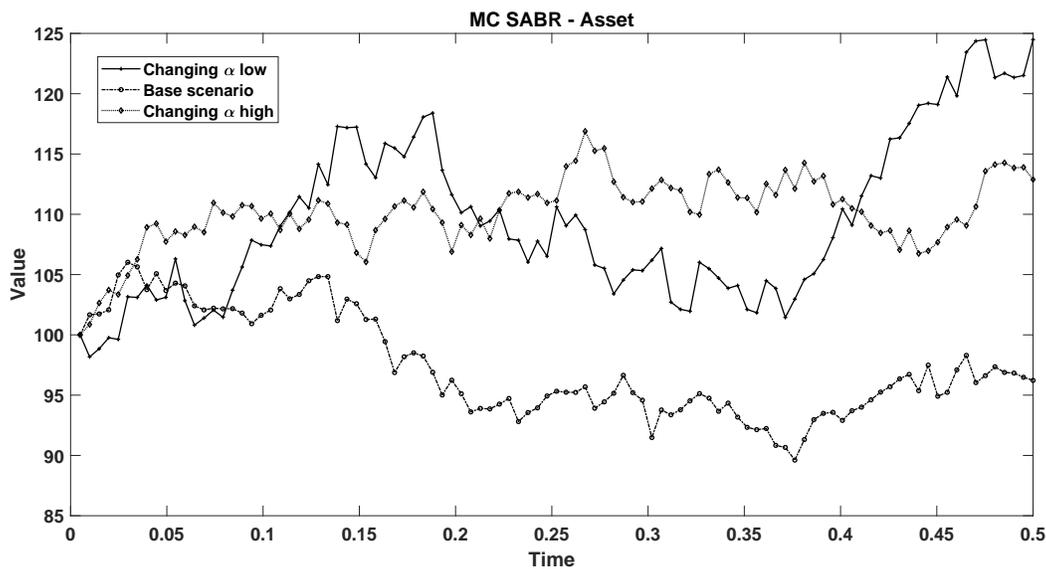}
   \caption{SABR - Asset}
   \label{Fig:SABR-V-alpha-c0} 
\end{subfigure}
\caption{$\alpha$ low = 0.1, $\alpha$ high = 0.5, $\alpha$ = 0.2}
\end{figure} 

\newpage
\pagebreak
\clearpage

\subsubsection{Sensitivity to $\alpha$, volatility paths}
To see how a path may change by changing the $\rho$ Fig. \ref{Fig:WN-V-alpha-c0} and Fig.  \ref{Fig:SABR-V-alpha-c0} show different paths of the volatility.
%
%
\begin{figure}[!ht]
    \centering
\begin{subfigure}[a]{1\textwidth}
   \includegraphics[width=1\linewidth]{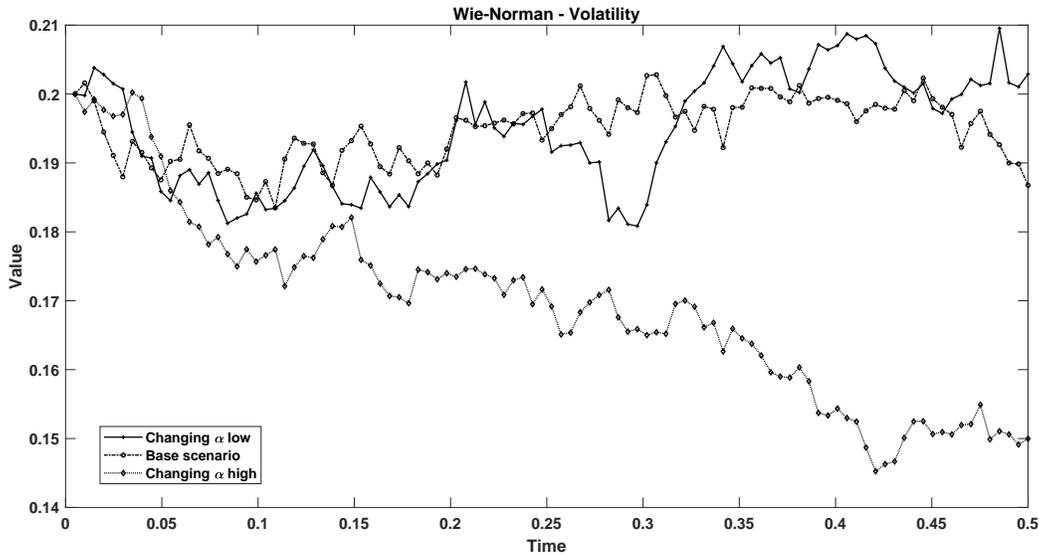}
   \caption{Wei-Norman SLV - Volatility}
   \label{Fig:WN-V-alpha-c0} 
\end{subfigure}
\begin{subfigure}[b]{1\textwidth}
   \includegraphics[width=1\linewidth]{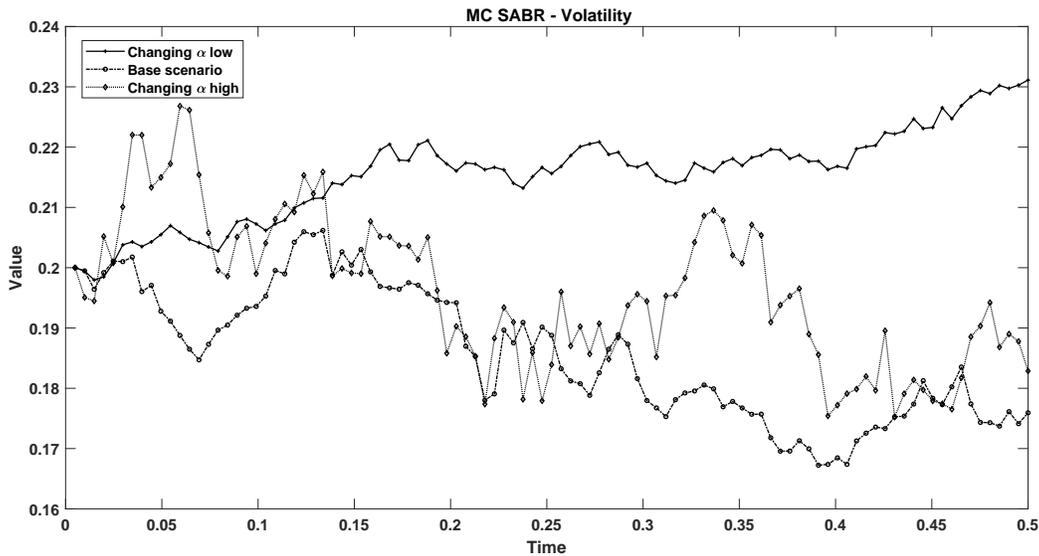}
   \caption{SABR - Volatility}
   \label{Fig:SABR-V-alpha-c0} 
\end{subfigure}
\caption{$\alpha$ low = 0.1, $\alpha$ high = 0.5, $\alpha$ = 0.2}
\end{figure} 

\subsection{Sensitivity to $\rho$}  \label{Sec:ParamRho}
This section shows how the two models perform when $\rho$ changes. As well as with the parameter $\alpha$, a higher value implies a higher trajectory for both $S$ and $v$. 

\subsubsection{Sensitivity to $\rho$, asset paths}

To see how a path may change by changing the $\rho$ Fig. \ref{Fig:WN-V-rho-c0} and Fig.  \ref{Fig:SABR-V-rho-c0} show different paths of the asset.
%
%
\begin{figure}[!ht]
    \centering
\begin{subfigure}[a]{1\textwidth}
   \includegraphics[width=1\linewidth]{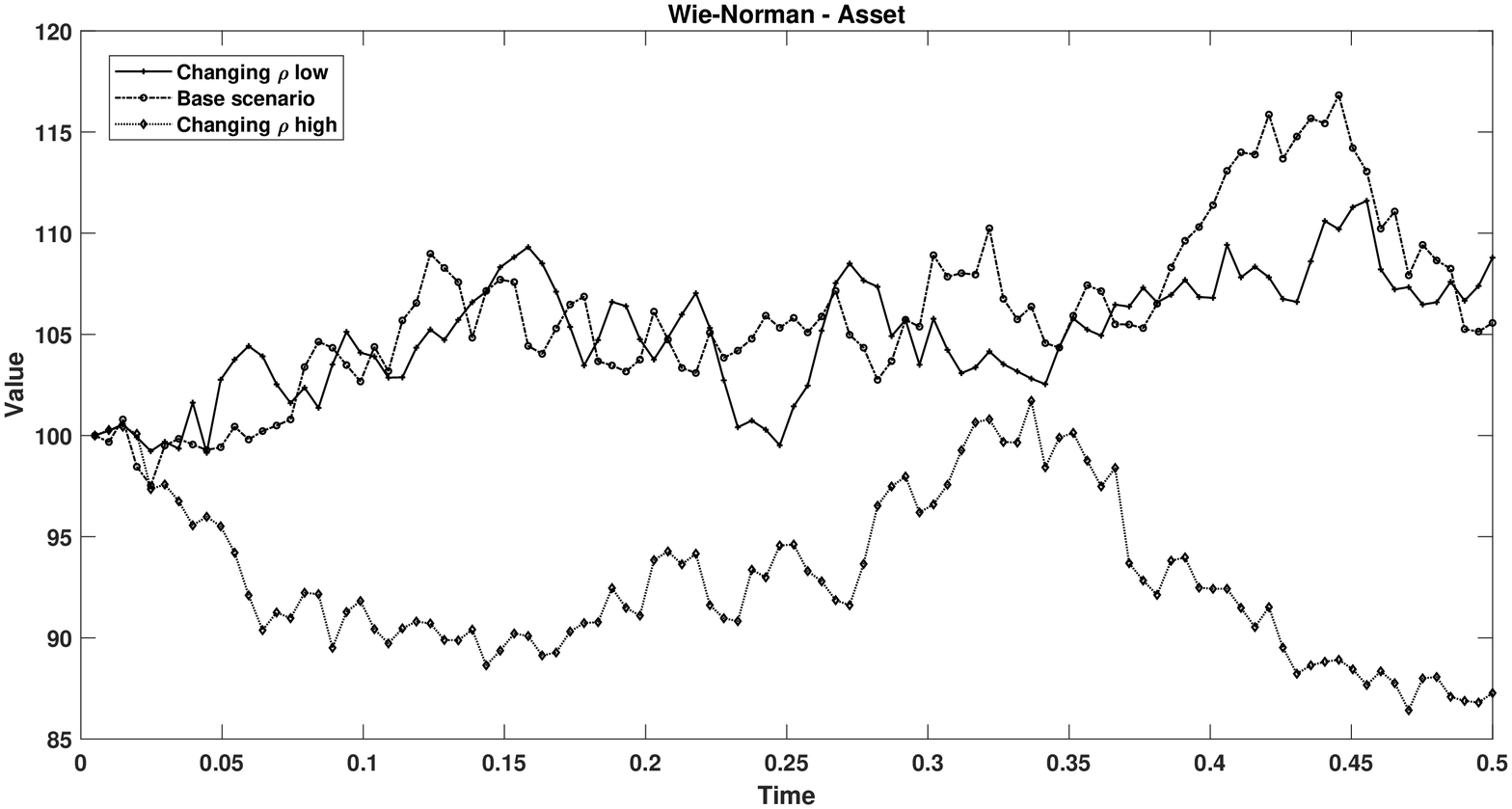}
   \caption{Wei-Norman SLV - Asset}
   \label{Fig:WN-V-rho-c0} 
\end{subfigure}
\begin{subfigure}[b]{1\textwidth}
   \includegraphics[width=1\linewidth]{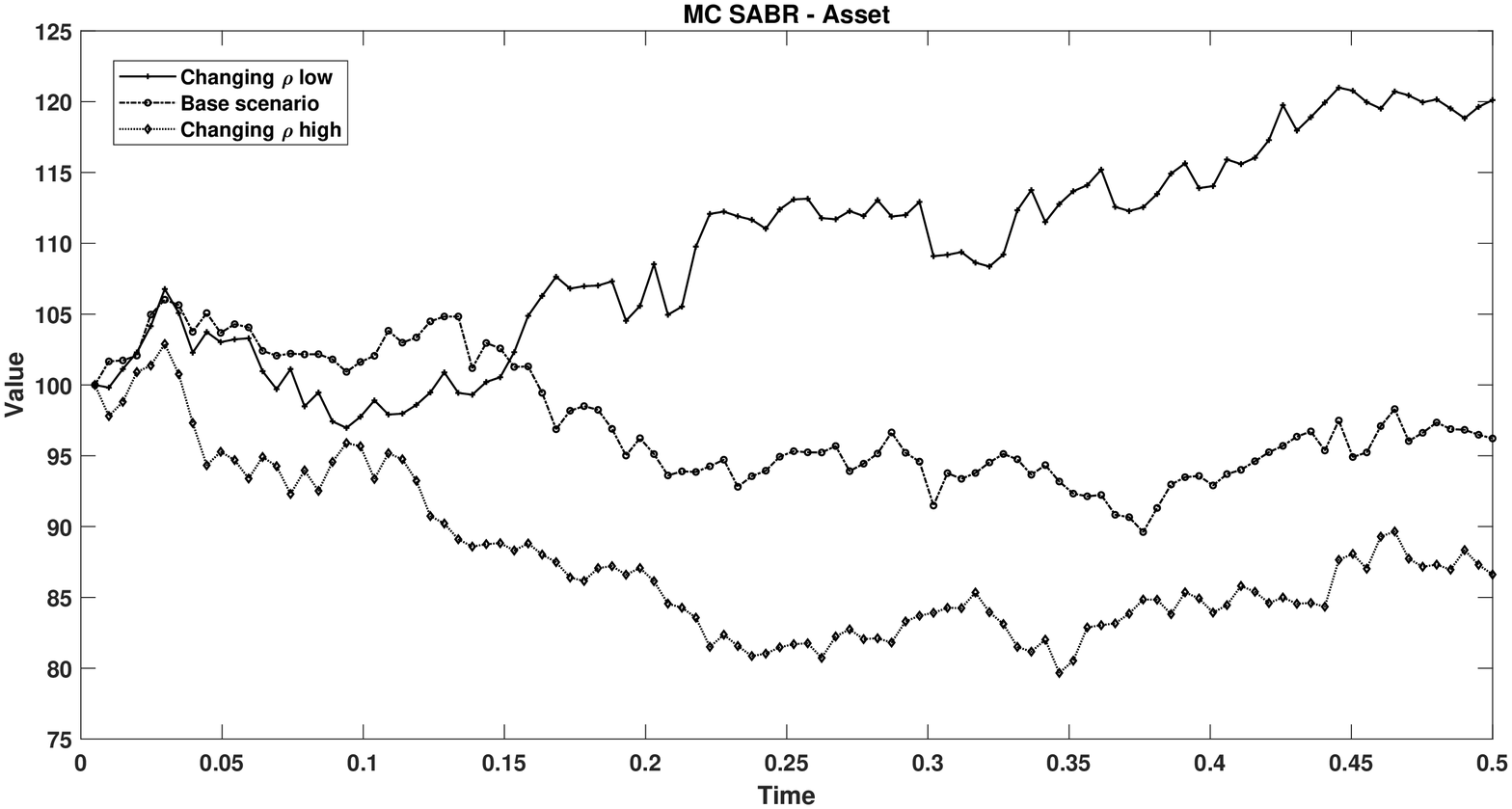}
   \caption{SABR - Asset}
   \label{Fig:SABR-V-rho-c0} 
\end{subfigure}
\caption{$\rho$ low = -0.8, $\rho$ high = 0.8, $\rho$ = 0.5}
\end{figure} 

\newpage
\pagebreak
\clearpage

\subsubsection{Sensitivity to $\rho$, volatility paths}
To see how a path may change by changing the $\rho$ Fig. \ref{Fig:WN-V-rho-c0} and Fig.  \ref{Fig:SABR-V-rho-c0} show different paths of the volatility.
%
%
\begin{figure}[!ht]
    \centering
\begin{subfigure}[a]{1\textwidth}
   \includegraphics[width=1\linewidth]{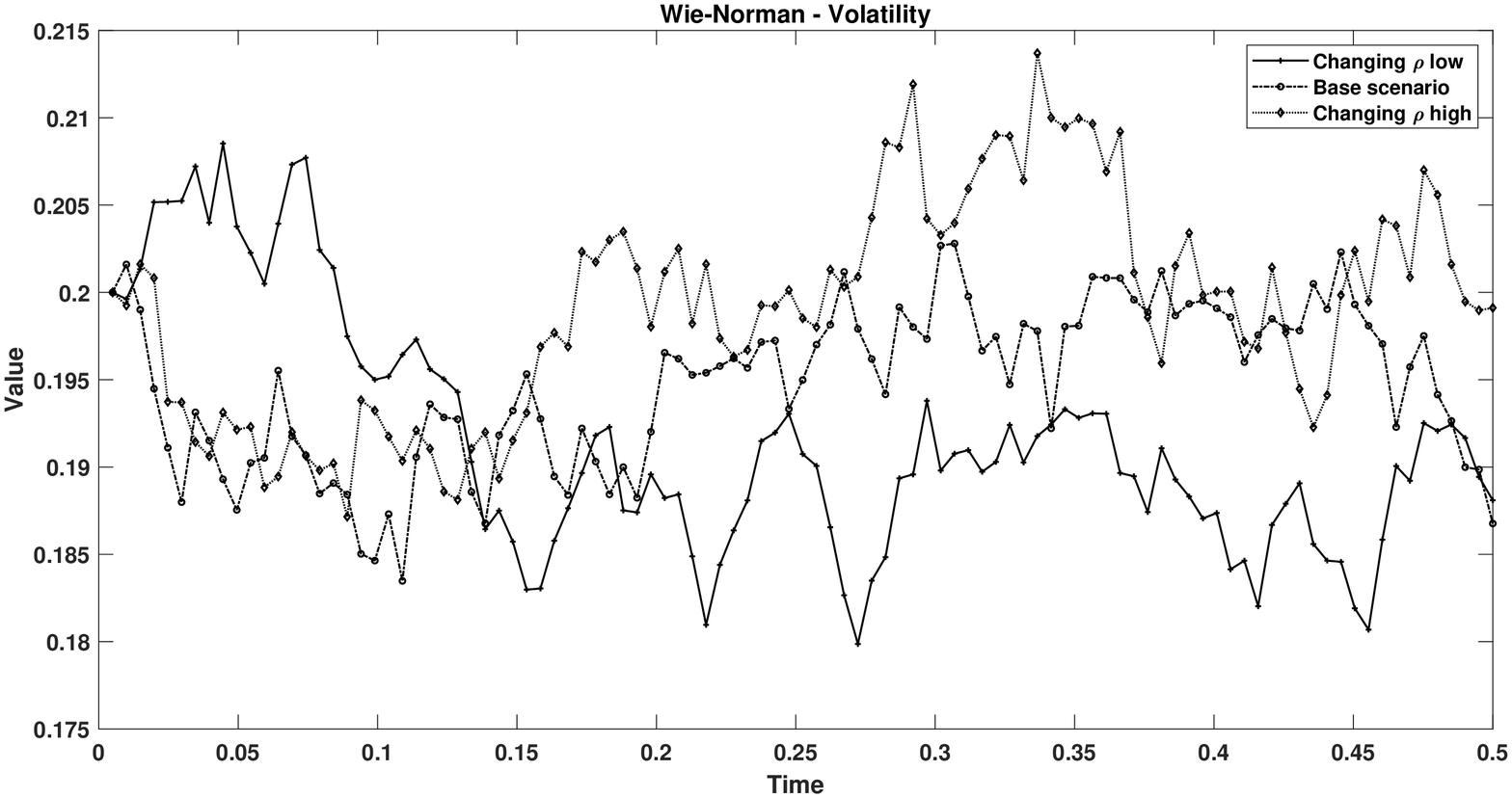}
   \caption{Wei-Norman SLV - Volatility}
   \label{Fig:WN-V-rho-c0} 
\end{subfigure}
\begin{subfigure}[b]{1\textwidth}
   \includegraphics[width=1\linewidth]{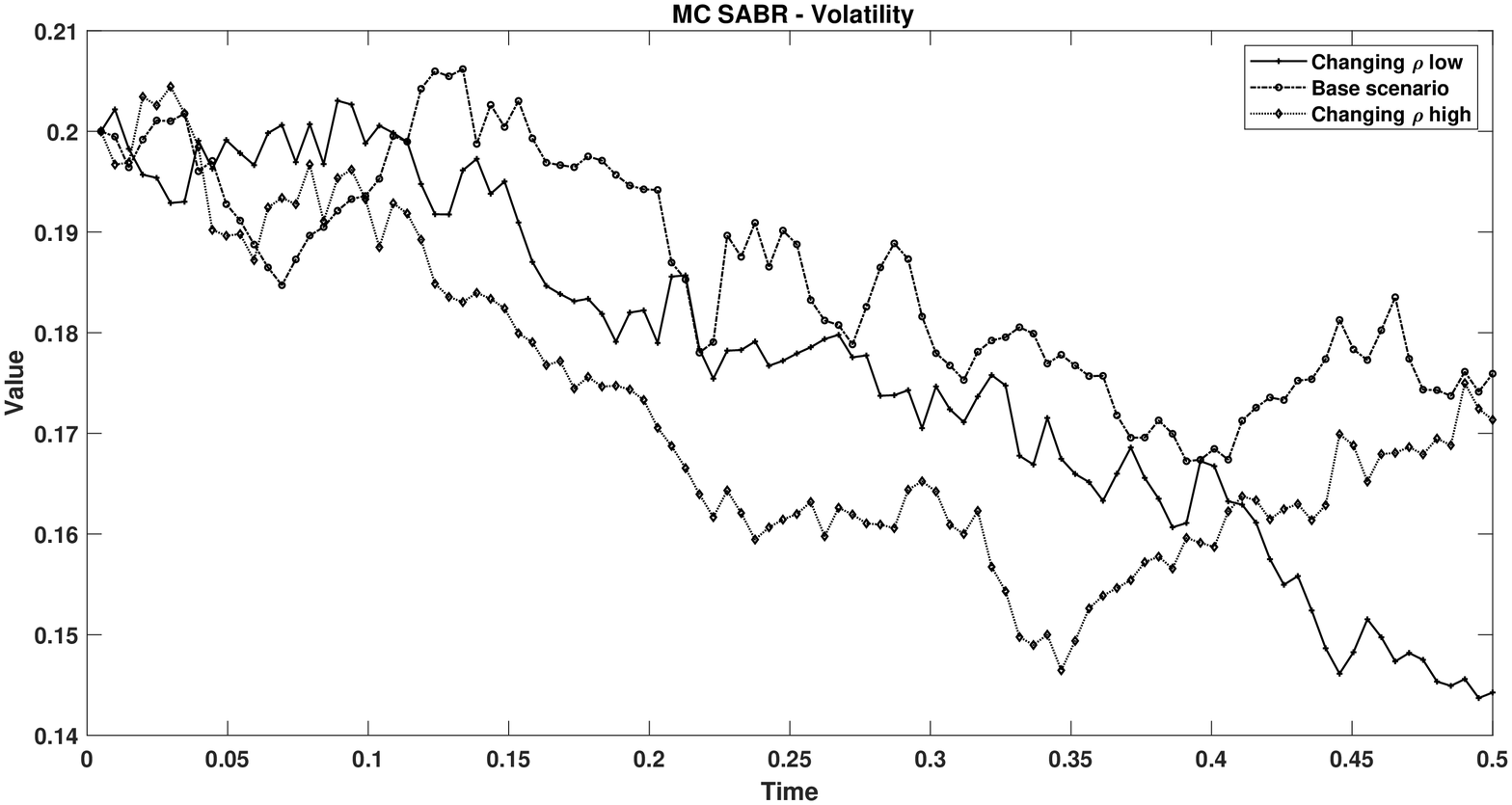}
   \caption{SABR - Volatility}
   \label{Fig:SABR-V-rho-c0} 
\end{subfigure}
\caption{$\rho$ low = -0.8, $\rho$ high = 0.8, $\rho$ = 0.5}
\end{figure} 

\newpage
\pagebreak
\clearpage

%
\bibliographystyle{siam} 

\bibliography{MyBibSLV}

\end{document}